\documentclass[12pt]{article}
\pdfoutput=1
\usepackage{amsmath,amssymb}
\usepackage{graphicx}
\usepackage{subfigure}
\usepackage[pdftex]{hyperref}
\DeclareGraphicsExtensions{.eps,.bmp,.wmf,.jpg,.pdf}
\numberwithin{equation}{section}
\def\be{\begin{equation}}
\def\ee{\end{equation}}

\def\bea{\begin{eqnarray}}
\def\eea{\end{eqnarray}}

\title{Natural scaling for dark energy}
\author{L.N. Granda\thanks{ngranda@univalle.edu.co, ngranda@um.es}\\{\small\it Departamento de Fisica, Universidad del Valle, A.A. 25360 Cali, Colombia}}

\date{}
\begin{document}
\maketitle

\begin{abstract}
\noindent We propose a dark energy density based on the Gauss-Bonnet 4-dimensional invariant and its modification. This model avoids the necessity of introducing the black hole limit to define the holographic density, since it can be considered as a non-saturated regime. This allows to describe the dark energy with an equation of state and Hubble parameter behaving in a way that can be adjusted very well to recent observations. The model presents quintom behavior without any future finite-time singularities. 
\end{abstract}

\section{Introduction}
\noindent The increasing amount of astrophysical data from distant Ia supernovae \cite{riess}, \cite{perlmutter}, \cite{hicken}, cosmic microwave background anisotropy \cite{komatsu}, \cite{larson}, and large scale galaxy surveys \cite{abazajian1}, \cite{tegmark3}, all indicate that the Universe is currently undergoing a phase of accelerated expansion. This late time acceleration is believed to be caused by some kind of  negative-pressure form of matter known as dark energy. The combined analysis of cosmological observations also suggests that the universe is spatially flat, and  consists of about $30\%$
of dark matter, and $70\%$ of homogeneously distributed dark energy with negative pressure. Despite this high percentage of the dark energy component, its nature as well as its cosmological origin remains unknown and represents one of the fundamental problems of theoretical cosmology. The cold dark matter model with a cosmological constant ($\Lambda$CDM) provides an excellent explanation for the accelerated expansion of the universe and other existing observational data \cite{jassal}, \cite{wilson}, \cite{davis}, \cite{allen}. Nevertheless according to the amount of astrophysical data, it remains very probably that the dark energy density is weakly time dependent, raising the interest for the search of a dynamical origin of dark energy. At the present a wide variety of models have been proposed to explain the nature of the dark energy and the accelerated expansion (see \cite{copeland, sergeiod} for review).
Among the different models of dark energy, the holographic dark energy approach is quite interesting as it incorporates some concepts of the quantum gravity known as the holographic principle (\cite{bekenstein, thooft, bousso, cohen, susskind}). According to the holographic principle, the entropy of a system scales not with its volume, but with its surface area. In the cosmological context, the holographic principle will set an upper bound on the entropy of the universe \cite{fischler}. In the work \cite{cohen}, a relationship between the short distance cut-off $\Lambda$ and the infra-red cut-off $L$ was suggested in the frame of quantum field theory. This relationship was established by using the limit set by black hole formation, namely, if is the quantum zero-point energy density caused by a short distance (UV) cut-off, the total energy in a region of size $L$ should not exceed the mass of a black hole of the same size, thus $L^3\Lambda^4\leq LM_p^2$. 
Applied to the dark energy issue, if we take the whole universe into account, then the vacuum energy related to this holographic principle is viewed as dark energy, usually called holographic dark energy \cite{cohen} \cite{hsu}, \cite{li}. The largest $L$ allowed is the one saturating this inequality so that the holographic dark energy density is defined by the equality $\rho_{\Lambda}=3c^2M_p^2L^{-2}$, where $c^2$ is a numerical constant and $M_p^{-2}=8\pi G$.\\
In the present work we propose a relation between the IR and UV cut-offs that obey the holographic bound $\rho_{\Lambda}\leq M_p^2L^{-2}$, without invoking the Planck scale. This relation can be established by proposing a dark energy density proportional to the Gauss-Bonnet (GB) 4-dimensional invariant and its modification. Besides his geometrical meaning, the GB invariant has the right dimension of energy density, and an energy density of the form $\rho\propto H^4$ can always obey the above mentioned bound for black hole formation. 
The main motivation for the present model is to consider the possibility of a dark energy density that scales in a natural way as $\rho_{\Lambda}\propto L^{-4}$ (i.e. with its volume), and respect the physical bounds set by the holographic principle (as follows from the relation $M_p^2>>H^2$).
As will be shown, the dark energy density proportional to the GB invariant or its modification leads to interesting cosmological consequences.  As we will see, this density gives the correct order of magnitude of the current critical density without involving the black hole bound, and could be considered as a non-saturated regime of the standard defined holographic principle.\\ 
The gauss-Bonnet term has been considered in many models of dark energy mainly as a term in the action, coupled to scalar field, or in modified theories including functions of the Gauss-Bonnet term $F({\cal G})$. The GB invariant is believed to give the leading-order correction to the low energy string gravity \cite{gross}, and its role on the dark energy issue was studied in \cite{odintsov5}, \cite{odintsov6}, \cite{odintsov7}, \cite{odintsov8}.
As it is well known, the infrared cut-off given by the future event horizon gives rise to accelerated expansion with an EoS parameter less than $-1/3$, but this model is based on non local quantities and faces problems with the causality \cite{li}. An holographic dark energy model which is based on local and non local quantities have been considered in \cite{sergei1, sergei2}. In \cite{gao}, \cite{granda}, \cite{granda1} an infrared cut-off for the holographic density was proposed, which describes the dark energy in good agreement with the astrophysical data, and may explain the cosmic coincidence.\\
The paper is organized as follows. In Sect. II we introduce the Gauss-Bonnet and Modified Gauss-Bonnet models and give a dynamical interpretation to the densities in terms of a tachyon scalar field. In sect. III we present an analysis of the cosmological dynamics, showing the evolution of the equation of state and the Hubble parameter. We also perform a statefinder diagnostic. Sect. IV is devoted to some discussion. 
\section{The model}
\subsection*{Gauss Bonnet}
Guided by facts as that the 4 dimensional Gauss-Bonnet invariant is quadratic in curvature (and therefore has the dimension of energy density), that it appears in quantum corrections to low energy string gravity \cite{gross}, and considering a non saturated regime in the holographic principle, we introduce the dark energy density as follows 
\begin{equation}\label{eq1}
\rho_\Lambda=\alpha{\cal G}
\ee
where $\alpha$ is a positive dimensionless parameter and ${\cal G}$ is the 4-dimensional Gauss-Bonnet invariant ${\cal G}=R^2-4R_{\mu\nu}R^{\mu\nu}+R_{\mu\nu\eta\gamma}R^{\mu\nu\eta\gamma}$. According to \cite{horvat}, this proposal could be considered as a lower bound for the holographic dark energy. It is worth noting that the renormalized vacuum density for scalar field with conformal invariance in de Sitter space time is proportional to $H^4$ \cite{dowker}, \cite{davies}.
The IR cut-off related with this density is given by the ``length'' size of the GB invariant $L\sim ({\cal G})^{-1/4}$. In the flat FRW background $ds^2=-dt^2+a(t)^2\sum^{3}_{i=1}(dx^i)^2$, the Eq. (\ref{eq1}) takes the form
\begin{equation}\label{eq2}
\rho_{\Lambda}=24\alpha\left(H^4+H^2\dot{H}\right)
\ee
In order to guarantee the observed current value of the DE density, $\rho_{DE}\sim M_p^2H_0^2$, it seems likely that we have to tune $\alpha$: according to (\ref{eq2}), $\rho_{\Lambda_0}\sim 24\alpha H_0^4$, which compared with $\rho_{DE}$ should give $24\alpha\sim M_p^2H_0^{-2}\sim 10^{122}$, but in fact the Friedmann equation together with the initial condition (i.e. the flatness condition in a flat FRW background) automatically take care about the correct magnitude of the dark energy density compatible with observations.\\
In absence of matter, the Friedmann equation with the energy density given by (\ref{eq2}), in the flat FRW background takes the form 
\be\label{eq3}
H^2=\frac{\kappa^2}{3}\rho_{\Lambda}=8\alpha \kappa^2 H^2\left(H^2+\dot{H}\right)
\ee
where $\kappa^2=8\pi G=M_p^{-2}$. After simplifying
\be\label{eq4}
\frac{dH}{dt}+H^2-\frac{1}{8\alpha\kappa^2}=0
\ee
This solution gives the asymptotic behavior of the Hubble parameter at late times. Similar equation has been obtained in \cite{capozziello} for a model of scalar field with non-minimal derivative couplings. The Eq. (\ref{eq4}) has the solution
\be\label{eq5}
H(t)=\frac{1}{(8\alpha\kappa^2)^{1/2}}\tanh\left[\frac{1}{\sqrt{8\alpha\kappa^2}}(t-t_0)\right]
\ee
This asymptotic solution describes a bouncing universe that at far future approaches the de Sitter solution $H|_{t\rightarrow \infty}=(\frac{1}{8\alpha\kappa^2})^{1/2}$. In \cite{easson}, this solution was also studied in a scalar-tensor model with auxiliary scalar field. In terms of the variable $x=\log a$ and the scaled Hubble parameter $\tilde{H}=H/H_0$, the Eq. (\ref{eq5}) becomes
\be\label{eq6}
\frac{1}{2}\frac{d\tilde{H}^2}{dx}+\tilde{H}^2-\tilde{A}=0
\ee
where $\tilde{A}=1/(8\alpha\kappa^2 H_0^2)$. Solving this equation with the initial condition $\tilde{H}^2|_{x=0}=1$, gives the solution
\be\label{eq7}
\tilde{H}^2=\tilde{A}+(1-\tilde{A}) e^{-2x}
\ee
Note that the second term behaves as $a^{-2}$, producing an effect similar to that of the spatial curvature. Note also that $\tilde{A}$ should satisfy $\tilde{A}\leq 1$, since otherwise it would give unphysical results (i.e. $\tilde{H}^2<0$ in the past). Let's draw some conclusions from the Eq. (\ref{eq7}) and the parameter $\tilde{A}$. If we assume that $\tilde{A}\sim 1$ (it can be compared with the parameter of the energy density associated with the cosmological constant, which is about 0.7), then from the definition of $\tilde{A}$ follows 
\be\label{eq8}
\alpha=\frac{M_p^2}{8\tilde{A}H_0^2}\approx\frac{M_p^2}{8H_0^2}
\ee
replacing back this value into the expression for the holographic density (\ref{eq2}), we give
\be\label{eq9}
\rho_{\Lambda}\approx 3M_p^2 H_0^{-2}\left(H^4+H^2\dot{H}\right)
\ee
giving the right magnitude of the current DE density $\rho_{\Lambda_0}\sim M_p^2 H_0^2$. So we don't need to involve from the beginning the Planck mass, and indeed the black hole bound, to obtain a probably interesting alternative for the holographic DE density. Compared with the usual saturated formula for the holographic density the present proposal could be interpreted as non saturated one. In this way the Friedmann equation and the initial condition take care about the current appropriate value for the dark energy density density.\\
The equation of state (EoS) parameter $w=-1-\frac{1}{3}\frac{d\tilde{H}^2/dx}{\tilde{H}^2}$, from (\ref{eq7}) is obtained as
\be\label{eq10}
w=-1+\frac{2}{3}\frac{(1-\tilde{A})e^{-2x}}{\tilde{A}+(1-\tilde{A} e^{-2x}}=-1+\frac{2}{3}\frac{(1-\tilde{A})(1+z)^2}{\tilde{A}+(1-\tilde{A})(1+z)^2}
\ee
in the last equality we used the redshift variable $1+z=e^{-x}$. This equation shows that $w<-1/3$, describing effectively evolving dark energy, with high redshift limit $w=-1/3$ at $z\rightarrow\infty$. At far future ($z\rightarrow -1$ or $x\rightarrow\infty$), the universe evolves toward de Sitter phase ($w\rightarrow-1$).
\subsection*{Modified Gauss Bonnet}
Let's consider the following modification of the GB dark energy density, that gives rise to interesting consequences. 
\be\label{eq11}
\rho_{\Lambda}=\gamma H^4+\delta H^2\dot{H}
\ee
where $\gamma$ and $\delta$ are dimensionless constants. Replacing this density in the Friedmann equation gives
\be\label{eq12}
\delta\frac{dH}{dt}+\gamma H^2-\frac{3}{\kappa^2}=0
\ee
Integrating this equation gives the solution of the type (\ref{eq5}) with the same asymptotic behavior. In the $x$ variable and using the scaled quantities, the Eq. (\ref{eq12}) becomes
\be\label{eq13}
\frac{1}{2}\frac{d\tilde{H}^2}{dx}+\frac{\gamma}{\delta}\tilde{H}^2-\frac{3}{\delta\kappa^2 H_0^2}=0
\ee
After integration with the initial condition $\tilde{H}^2|_{x=0}=1$, leads to the solution 
\be\label{eq14}
\tilde{H}^2=\tilde{B}+(1-\tilde{B}) e^{-\frac{2\gamma}{\delta}x}
\ee
where $\tilde{B}=3/(\gamma\kappa^2 H_0^2)$. An important characteristic of this solution is that coefficient of the exponent depends on the constants $\gamma$ and $\delta$, while the first term (corresponding to some density parameter) depends only on one constant $\gamma$, so we can handle them independently. According to this solution, if we consider $\gamma/\delta<0$, then the phantom behavior for dark energy is possible. In fact we can adjust these constants with the available observational data. For the same reasons discussed above, we can consider $\tilde{B}\sim 1$, which again reproduces the correct expression for the observed DE density $\rho_{\Lambda_0}\sim M_p^2 H_0^2$. The dark energy EoS parameter from (\ref{eq14}) is given by
\be\label{eq15}
w=-1+\frac{2\gamma}{3\delta}\frac{(1-\tilde{B}) e^{-2(\gamma/\delta)x}}{\tilde{B}+(1-\tilde{B}) e^{-2(\gamma/\delta)x}}=-1+\frac{2\gamma}{3\delta}\frac{(1-\tilde{B}) (1+z)^{2(\gamma/\delta)}}{\tilde{B}+(1-\tilde{B}) (1+z)^{2(\gamma/\delta)}}
\ee
for $\delta>0$ ($\gamma$ must always be positive), this EoS parameter runs between $w=-1+\frac{2\gamma}{3\delta}$ at $z\rightarrow \infty$ and $w=-1$ at $z\rightarrow -1$ (evolving towards de Sitter phase at future). Note that for $\gamma/\delta=3/2$ the EoS describes presureless matter-like behavior at  $z\rightarrow \infty$.\\
Compared to the holographic model proposed in \cite{granda},\cite{granda1} in absence of matter, in that case we obtained power-law solution giving rise to constant EoS (see \cite{granda1}), and in the present model we obtained that even in absence of matter the EoS evolves dynamically.
\subsection*{Dynamical Interpretation}
In the present case, when we consider only the presence of dark energy fluid with density given by (\ref{eq2}) or (\ref{eq11}), we can find an exact expression for the scalar field and potential corresponding to this fluid.\\
Let's consider the scalar tachyon field with the action
\be\label{eq11a}
S=-\int d^4x V(\phi)\sqrt{-\det\left[g_{\mu\nu}+\partial_{\mu}\phi\partial_{\nu}\phi\right]}
\ee
In the flat Friedmann background, the energy density and pressure density of the tachyon scalar field are given by \cite{padmanabhan, copeland}
\be\label{eq11b}
\rho_{\phi}=\frac{V(\phi)}{\sqrt{1-\dot{\phi}^2}},\,\,\,\, p_{\phi}=-V(\phi)\sqrt{1-\dot{\phi}^2}
\ee
And the Einstein's equations take the form
\be\label{eq11c}
H^2=\frac{\kappa^2}{3}\frac{V(\phi)}{\sqrt{1-\dot{\phi}^2}}
\ee
and
\be\label{eq11d}
3H^2+2\dot{H}=\kappa^2 V(\phi)\sqrt{1-\dot{\phi}^2}
\ee
Combining Eqs. (\ref{eq11c}) and (\ref{eq11d}) one can express the tachyon field and the potential as follows
\be\label{eq11e}
\dot{\phi}^2=-\frac{2}{3}\frac{\dot{H}}{H^2},
\ee
and 
\be\label{eq11f}
V(\phi)=\frac{3H^2}{\kappa^2}\left[1+\frac{2\dot{H}}{3H^2}\right]^{1/2}
\ee
Using the solution (\ref{eq14}) and solving the Eq. (\ref{eq11e}) in the $x$ variable, one finds
\be\label{eq11g}
\phi=\sqrt{\frac{2\delta}{3\gamma \tilde{B}}}\frac{1}{H_0}\arctan\left[\sqrt{\frac{\tilde{B}}{1-\tilde{B}}}e^{\gamma x/\delta}\right]
\ee
where $\delta>0$. And from (\ref{eq11f}) one can find the potential in terms of the scalar field as
\be\label{eq11h}
V(\phi)=\frac{3\tilde{B}H_0^2}{\kappa^2}\frac{1}{\tan^2(\tilde{\phi})}\left[3\tan^4(\tilde{\phi})+2(3-\frac{\gamma}{\delta})\tan^2(\tilde{\phi})+3-\frac{2\gamma}{\delta}\right]^{1/2}
\ee
where $\tilde{\phi}=\sqrt{\frac{3\gamma\tilde{B}}{2\delta}}\phi$. The GB case is obtained for $\gamma=\delta$. Is worth noting that the usual quintessence scalar field leads to inconsistencies (negative potential or complex $\phi$ for $x>0$) when trying to represent the solution (\ref{eq7}) or (\ref{eq14}) for $\delta>0$. So, the dynamics generated by the dark energy density (\ref{eq2}) or (\ref{eq11}) can be represented by an scalar field of the tachyon type with the potential (\ref{eq11h}).\\

\section{Adding matter content}
Here we consider the complete model with the matter component (usual barionic and dark matter). As we will see, adding the matter component causes strong changes in the differential equation, which becomes non-linear, and numerical calculation shows interesting results that deserve study. We will consider the general case of modified Gauss Bonnet model (MGB) and then analyze different cases, particularly the Gauss Bonnet model. 
Adding the matter term $\rho_m=\rho_{m0}e^{-3x}$ to the Friedmann equation gives
\be\label{eq16}
H^2=\frac{\kappa^2}{3}\left(\gamma H^4+\delta H^2\dot{H}+\rho_{m0} e^{-3x}\right)
\ee
which in the $x$ variable and using the scaled magnitudes, takes the form
\be\label{eq17}
\frac{\tilde{\delta}}{2}\tilde{H}^2\frac{d\tilde{H}^2}{dx}+\tilde{\gamma}\tilde{H}^4-\tilde{H}^2+\Omega_{m0}e^{-3x}=0
\ee
where $\tilde{\gamma}=\kappa^2H_0^2\gamma/3$, $\tilde{\delta}=\kappa^2H_0^2\delta/3$ and $\Omega_{m0}=\kappa^2\rho_{m0}/(3H_0^2)$ ($\Omega_m=\kappa^2\rho_m/(3H^2)$). This equation should be solved with the initial condition
\be\label{eq18}
\frac{d\tilde{H}^2}{dx}\Big|_{x=0}=1
\ee
Note that this equation is more complicated that (\ref{eq4}) because in absence of matter we could cancel one $H^2$ factor, which considerably simplified the equation. This equation can not be solved analytically, but can be integrated numerically in a given redshift interval and for given values of the parameters $\tilde{\gamma}$ and $\tilde{\delta}$, taking into account the initial (flatness) condition (\ref{eq18}). Let's consider first the GB case which corresponds to $\tilde{\gamma}=\tilde{\delta}=8\kappa^2H_0^2\alpha=\tilde{\alpha}$ (see (\ref{eq3})).
Solving numerically (\ref{eq17}) in the redshift variable with $\Omega_{m0}=0.31$ and the initial condition (\ref{eq18}), we obtain the behavior of the EoS parameter as described in Fig. 1, for two values of $\alpha$.
\begin{center}
\includegraphics [scale=0.7]{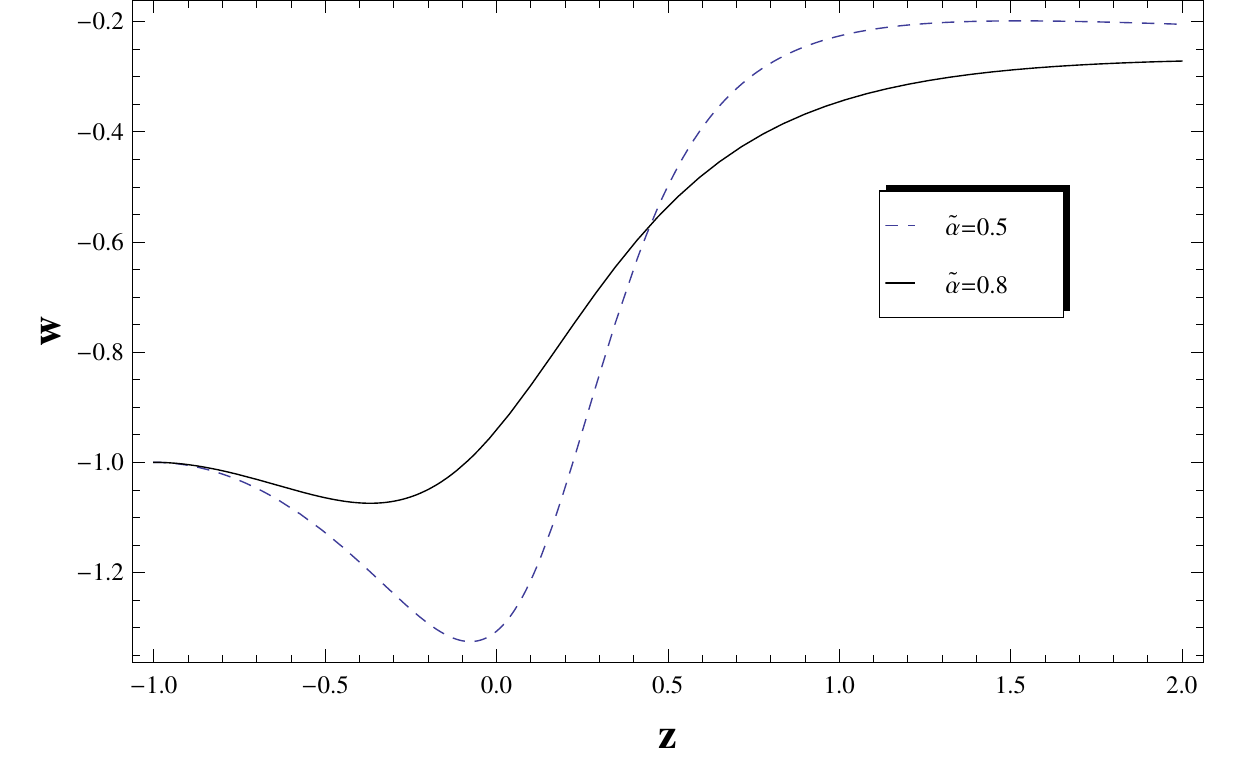}
\end{center}
\begin{center}
{fig. 1 \it The EoS parameter for the GB DE density  with $\Omega_{m0}=0.31$, $\tilde{\alpha}=0.55$ (dashed line) and $\tilde{\alpha}=0.8$.}
\end{center}
And in Fig. 2 we show the evolution of $H(z)$ for the corresponding parameters of Fig. 1 against its observational values with error bars \cite{simon, stern}.\\
\begin{center}
\includegraphics [scale=0.7]{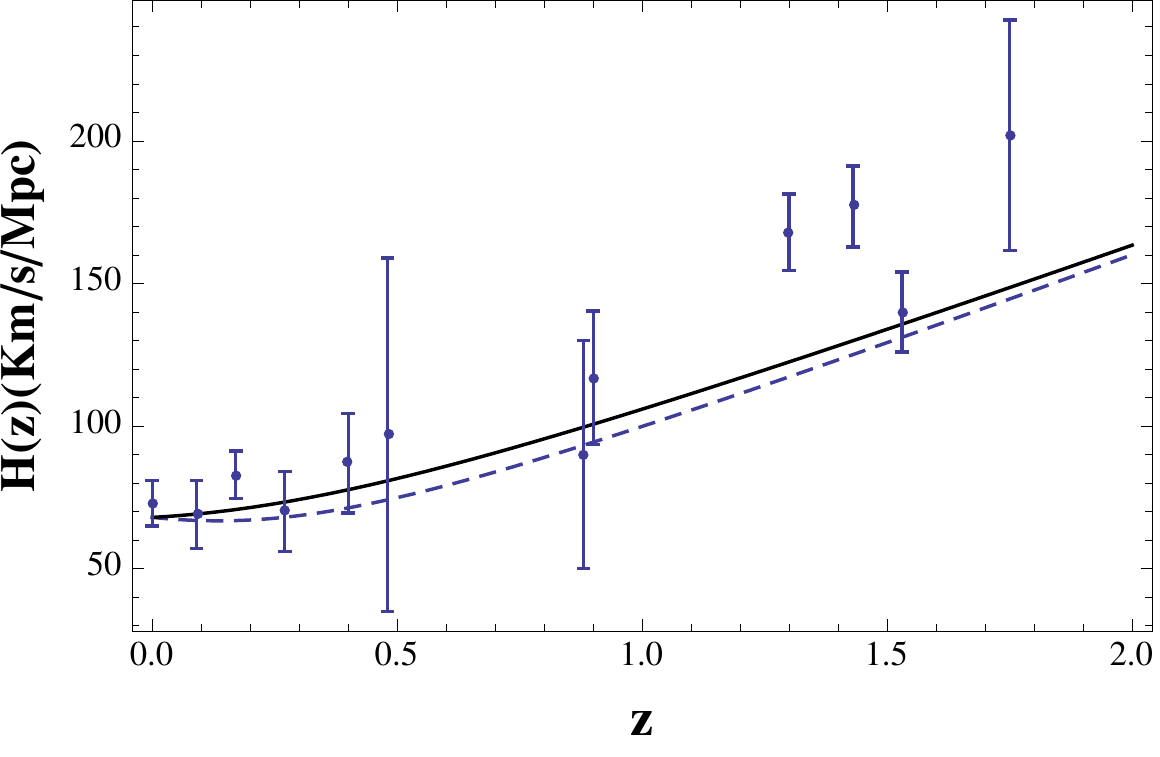}
\end{center}
\begin{center}
{Fig. 2 \it The observational $H(z)$ data with error bars and $H(z)$ form the solution of Eq. (\ref{eq17}) (for $\gamma=\delta$) with $\Omega_{m0}=0.31$, $\tilde{\alpha}=0.55$ (dashed line) and $\tilde{\alpha}=0.8$.}
\end{center}
The EoS parameter has been calculated according to
\be\label{eq19}
w=-1-\frac{2}{3}\frac{\dot{H}}{H^2}=-1+\frac{1+z}{3\tilde{H}^2}\frac{d\tilde{H}^2}{dz}
\ee
In the case $\tilde{\alpha}=0.8$, the current EoS parameter takes the value $w_0\sim -0.9$, and for $\tilde{\alpha}=0.55$ we have $w_0\sim -1.17$, so the model allows for quintom behavior. In both cases the EoS has a minimum value at future in the phantom region $w_{min}<-1$, and then evolves toward de Sitter phase.\\
For the MGB model, after numerical integration of Eq. (\ref{eq17}) with the initial condition (\ref{eq18}), $\Omega_{m0}=0.27$ and different choices of $\tilde{\gamma}$ and $\tilde{\delta}$ we found the EoS as showed in Fig. 3
\begin{center}
\includegraphics [scale=0.7]{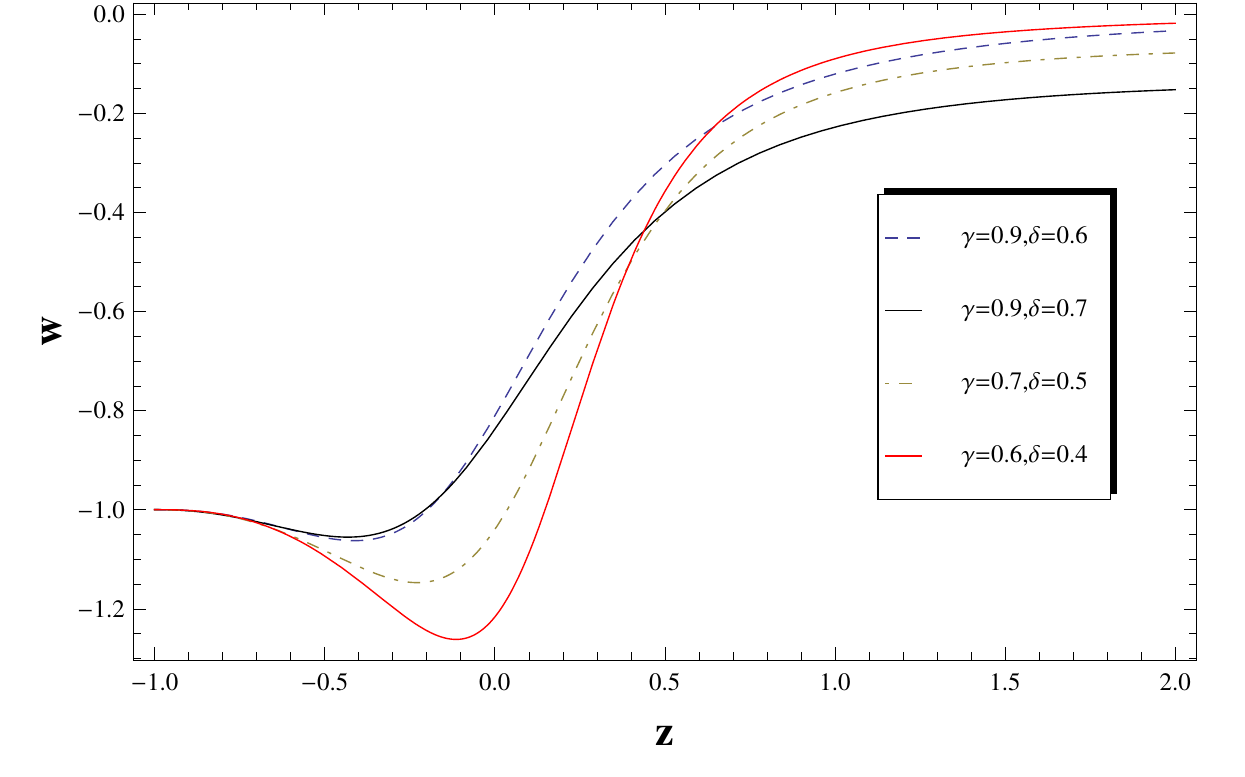}
\end{center}
\begin{center}
{Fig. 3 \it The EoS parameter for the MGB DE density  with $\Omega_{m0}=0.31$, and some representative values for the parameters $(\tilde{\gamma},\tilde{\delta})$ as shown in the graphic.}
\end{center}
In Fig. 4 we show evolution of $H(z)$ for the corresponding parameters of Fig. 3 against the observational values with error bars.
\begin{center}
\includegraphics [scale=0.7]{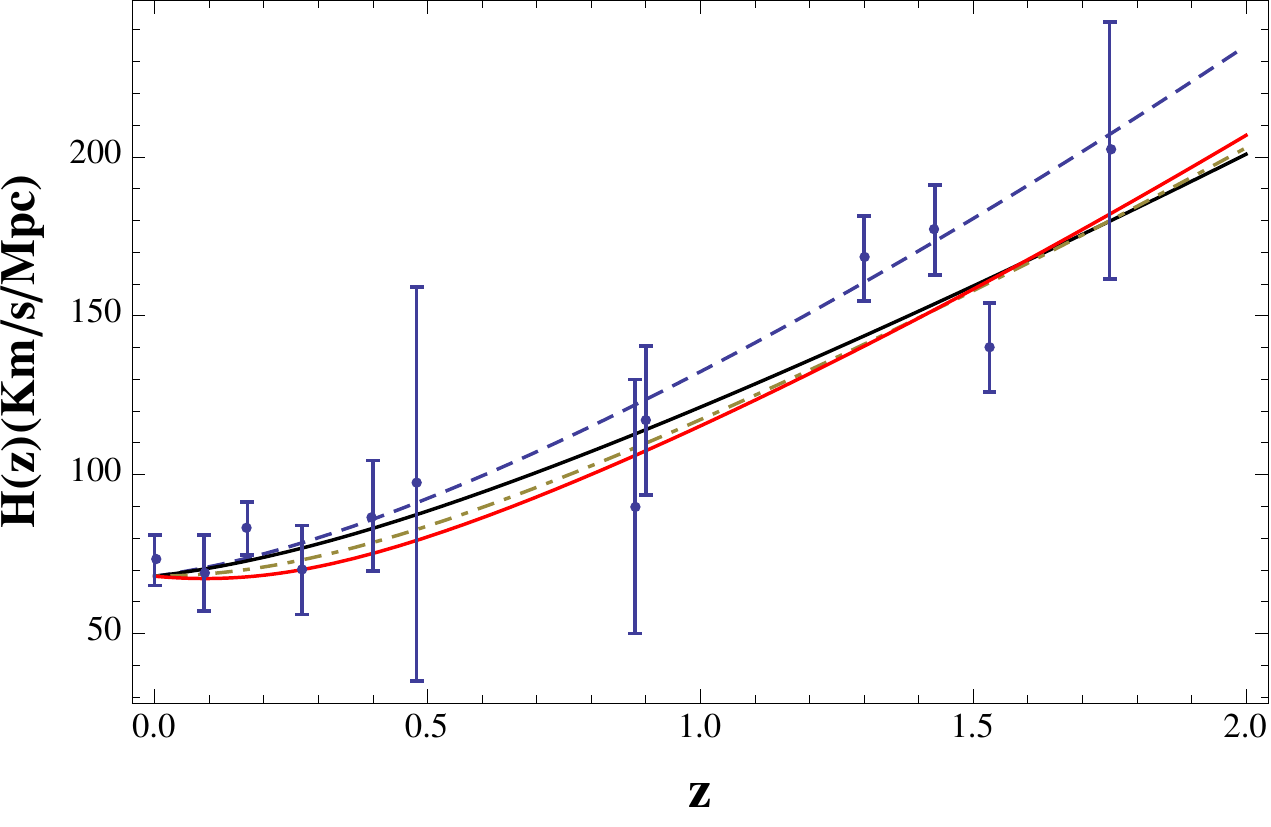}
\end{center}
\begin{center}
{Fig. 4 \it The observational $H(z)$ data with error bars, and $H(z)$ from the solution of Eq. (\ref{eq17}) for the corresponding curves of Fig. 3}
\end{center}
Corresponding to each pair of parameters $(\tilde{\gamma},\tilde{\delta})$, the current values for EoS parameter are: $w_0\sim -0.77$ for $(0.9,0.6)$, $w_0\sim -0.8$ for $(0.9,0.7)$, $w_0\sim -0.99$ for $(0.7,0.5)$ and $w_0\sim -1.15$ for $(0.6,0.4)$. The last point gives currently quintom behavior, but all curves exhibit a minimum at future, bellow the phantom divide and then evolve towards the de Sitter phase. In all this analysis we have guided by the last data provided by the Planck Collaboration \cite{planck1,planck2}, namely $\Omega_m\sim 0.31$ and $H_0\sim 68 Km/s/Mpc$.
\subsection*{Statefinder Diagnostic}
An useful tool to compare different dark energy models uses derivatives of the scale factor beyond the second order \cite{sahni1, sahni2}. The diagnostic proposal called ``statefinder'' introduces new geometrical dimensionless parameters that characterize the properties of dark energy regardless of the model, as they depend on the observable Hubble parameter and its derivatives. The statefinder parameters $q$, $r$ and $s$ ($q$ is also known as the deceleration parameter) are defined as
\be\label{eq20}
q=-1-\frac{\dot{H}}{H^2},\,\,\, r=\frac{\dddot{a}}{aH^3},\,\,\,\, s=\frac{r-1}{3(q-1/2)}
\ee
For our analysis will be useful to write these parameters in terms of the Hubble parameter $H(z)$ and the redshift $z$ as 
\be\label{eq21}
\begin{aligned}
q=-1+\frac{1+z}{2H^2}\frac{dH^2}{dz}&,\,\,\, r=1+\frac{1}{2H^2}\left[(1+z)^2\frac{d^2H^2}{dz^2}-2(1+z)\frac{dH^2}{dz^2}\right]\\
& s=\frac{1}{3}\frac{(1+z)^2\frac{d^2H^2}{dz^2}-2(1+z)\frac{dH^2}{dz}}{(1+z)\frac{dH^2}{dz}-3H^2}
\end{aligned}
\ee
An important characteristic of the pair ($r,s$) is that the spatially flat $\Lambda$CDM scenario corresponds to a fixed point ($1,0$) in the $s-r$ plane, with respect to which we can contrast the trajectories of other dark energy models. Though the relations between these statefinder parameters, namely $r(s)$ and
$r(q)$ can not be derived analytically for the present model, we give here a numerical analysis and plot a representative statefinder diagrams for the GB and the MGB models. In Fig. 5 we show the statefinder trajectories in the $s-r$ and $q-r$ planes for the GB model, taking $\tilde{\alpha}=0.8$ and Fig. 6 shows the statefinder trajectories for the MGB, taking $\tilde{\gamma}=0.9$ and $\tilde{\delta}=0.6$.\\ 
\begin{figure}[hbtp]
\begin{center}
\includegraphics [scale=0.55]{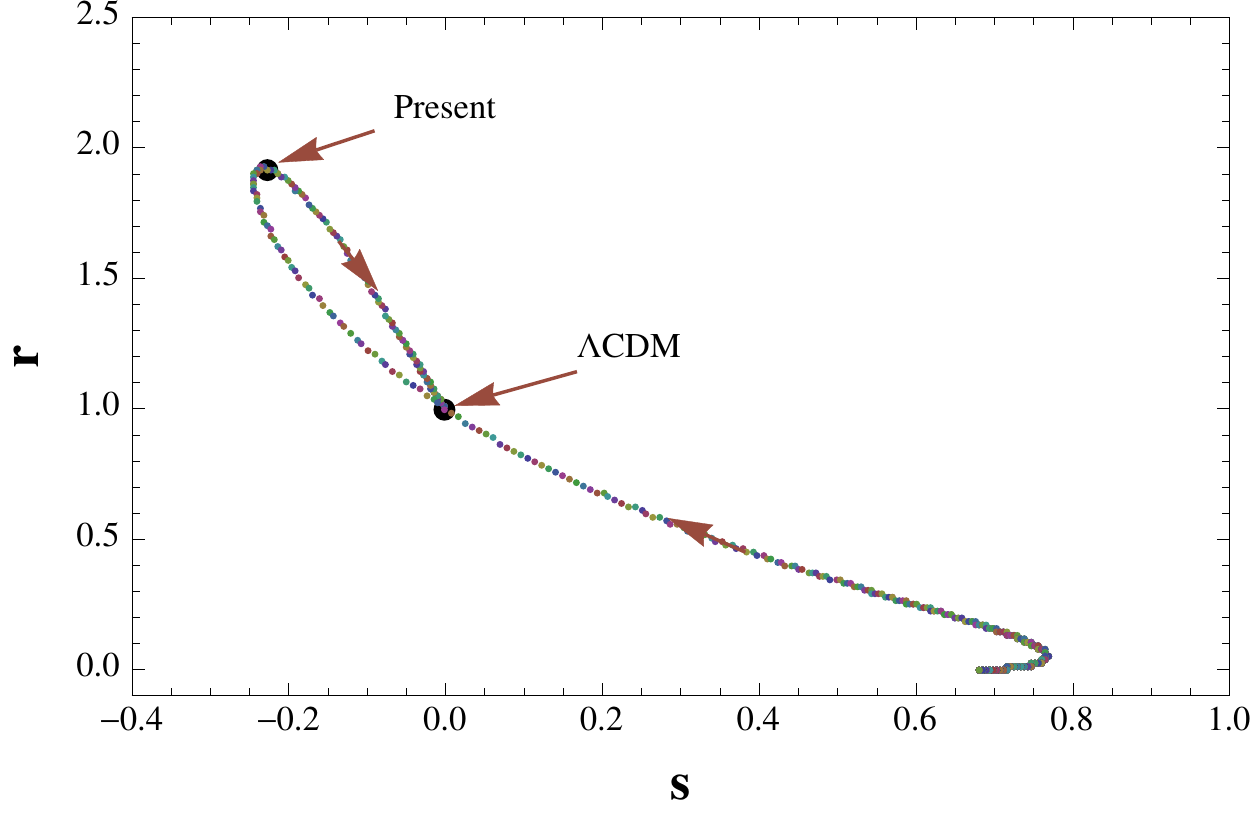}
\includegraphics [scale=0.55]{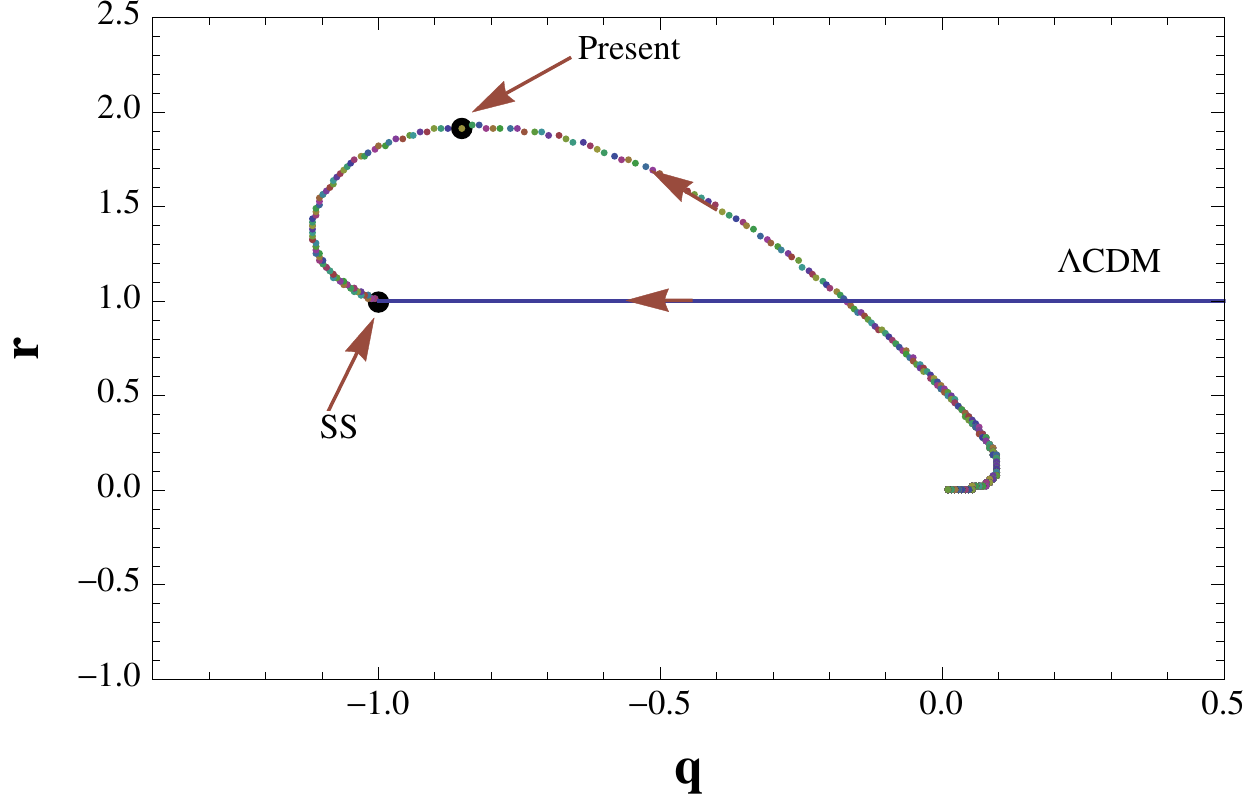}
\end{center}
\begin{center}
{Fig. 5. \it Evolution of the universe in the statefinder $s-r$ plane (left) and $q-r$ plane (right) for the GB DE model with $\tilde{\alpha}=0.8$, where the arrows along the curves denote the direction of evolution.}
\end{center}
\end{figure}
\noindent In the left graphic in Fig. 5, the pair ($r,s$) starts at the right of the $\Lambda$CDM fixed point, which is characteristic of quintessence behavior ($0<s<1$, $r<1$) and then evolves to the left of the $\Lambda$CDM (characteristic region of the Chaplygin gas model) \cite{sahni2} to asymptotically approach again the $\Lambda$CDM at late times. The trajectory in the $r-q$ plane starts in the region ($0<s<1$, $r<1$) and then crosses the $\Lambda$CDM line at some redshift in the past to evolve towards the de Sitter expansion at the future ($\Lambda$CDM$\rightarrow SS$ at $t\rightarrow \infty$). The trajectory $r(q)$ reaches a turning point in the phantom region ($q<-1$) and then evolves back to the SS point. The current values of the statefinder parameters are: $r_0\approx 1.92, s_0\approx -0.22, q_0\approx -0.86$.\\
\begin{figure}[hbtp]
\begin{center}
\includegraphics [scale=0.55]{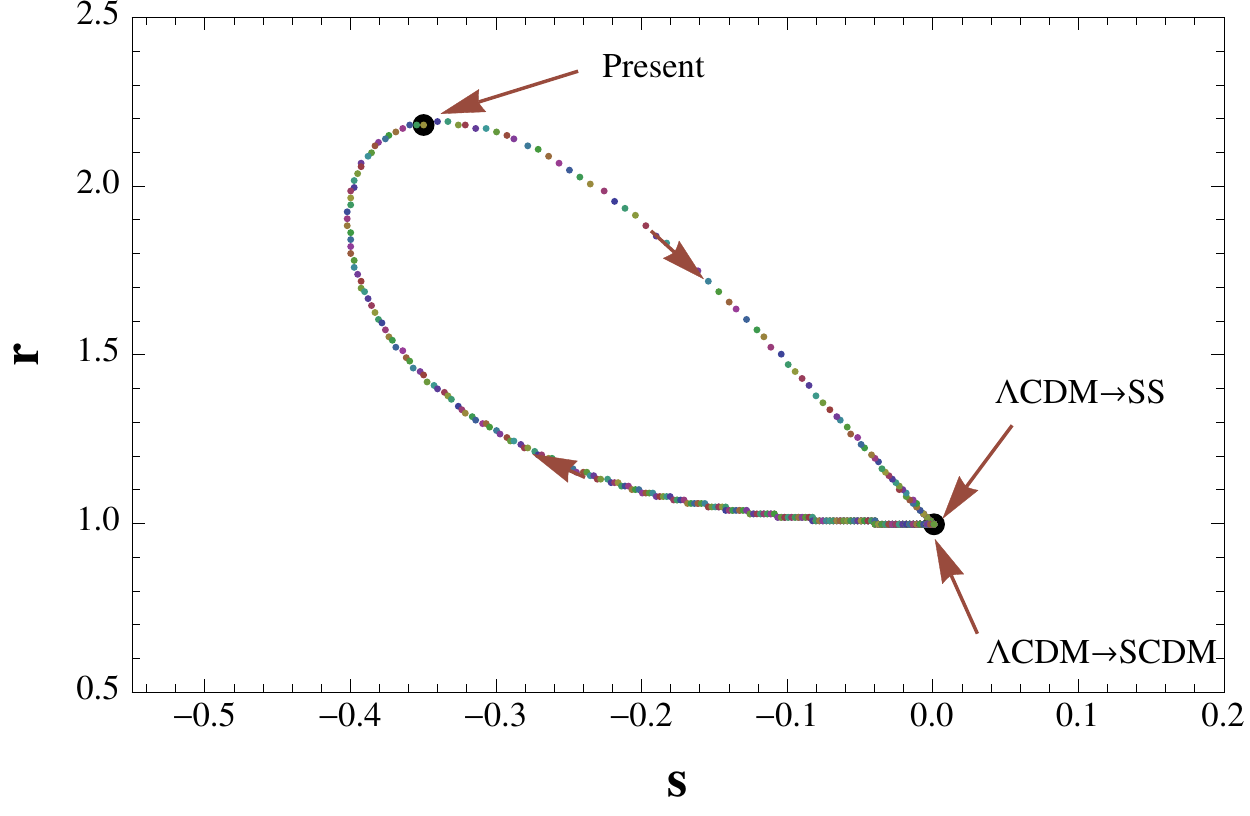}
\includegraphics [scale=0.55]{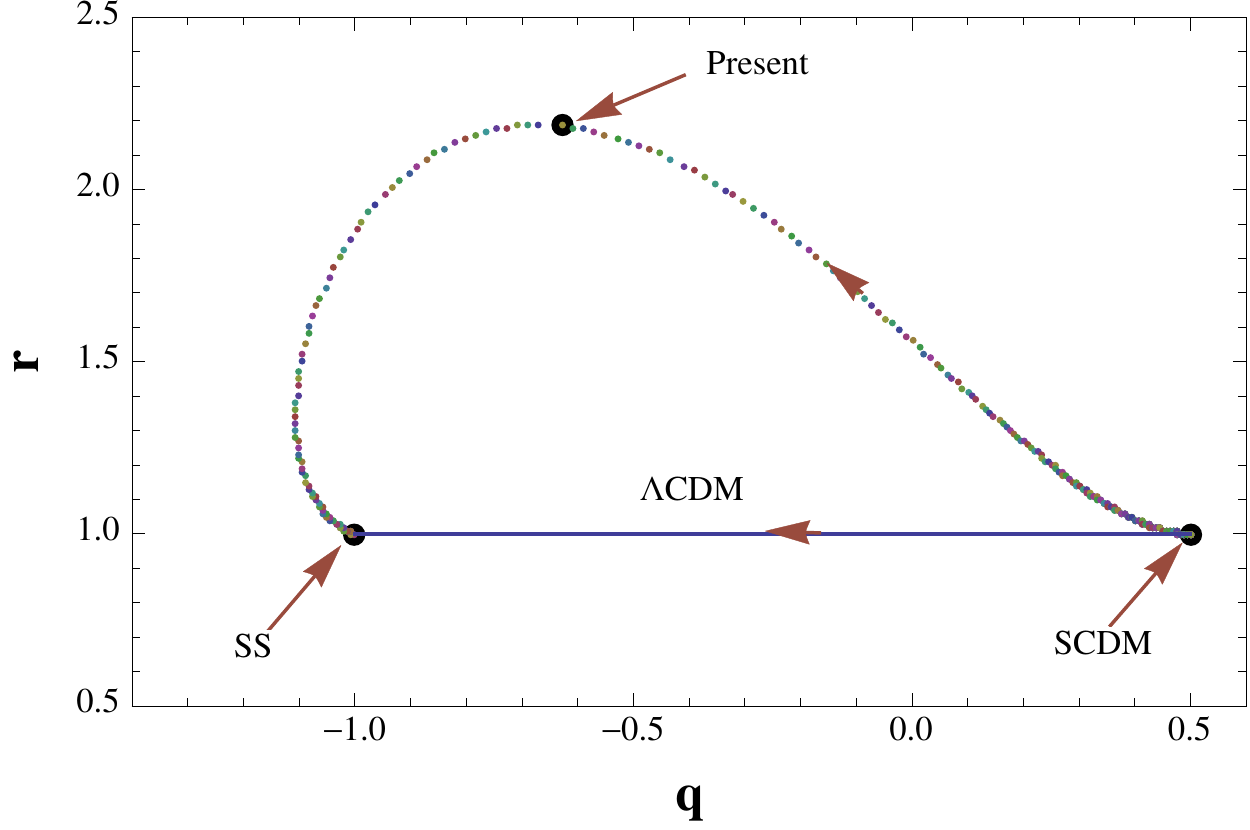}
\end{center}
\begin{center}
{Fig. 6. \it Evolution of the universe in the statefinder $s-r$ plane (left) and $q-r$ plane (right) for the MGB DE model with $\tilde{\gamma}=0.9$ and $\tilde{\delta}=0.6$.}
\end{center}
\end{figure}
In Fig. 6 the parameter $r$ increases monotonically from unity to a maximum value and then decreases to unity. In this case the $r(s)$ trajectory approaches the steady state model (SS) (the de Sitter phase) asymptotically at late times ($\Lambda$CDM$\rightarrow SS$ at $t\rightarrow\infty$, $\Omega_m\rightarrow 0$). The evolution in the $r-q$ plane takes place in the upper half (over the $\Lambda$CDM line) and clearly shows that the evolution starts at the SCDM and ends at the SS model. The parameter $q$ reaches the minimum value in the future, in the region $q<-1$ (phantom phase) and turns back asymptotically to the de sitter expansion. The current values for the statefinder parameters are: $r_0\approx 2.19, s_0\approx -0.34, q_0\approx -0.65$.\\
According to the trajectories depicted above for both models, the expansion crosses the phantom divide at some point in the future, reaching a turning point in the region $q<-1$ and then evolving toward the de Sitter phase. The current value of $r$ is very close to the maximum value for both models. The $r(s)$ trajectory for the MGB model is a closed loop that starts in the early time limit of the $\Lambda$CDM fixed point (SCDM) and ends in the $t\rightarrow \infty$ limit of the $\Lambda$CDM (SS) with de Sitter expansion.\\
\section{Discussion}
We proposed an infrared cut-off for the DE density, based on the linear combination of $H^4$ and $H^2\dot{H}$ terms, that for particular values of the coefficients gives the GB 4-dimensional topological invariant. This proposal has the right dimension of density without introducing dimensional parameters and may be interpreted as a non-saturated regime in the conventional holographic principle, with the advantage that we don't need to resort to the limit imposed by the black hole formation. After solving the Friedamnn equation with the initial condition we get an expression for the dark energy density that incorporates the Planck mass, and behaves in a way compatible with observations. We have found a scalar tachyon field with potential given by Eq. (\ref{eq11h}), that represents the solutions (\ref{eq7}) and (\ref{eq14}), giving a dynamical interpretation to the models (\ref{eq2}) and (\ref{eq11}).
Unlike the holographic density proposed in \cite{granda}, \cite{granda1} in the absence of matter, the present model describes a dynamical varying EoS as stated by Eqs. (\ref{eq10}) and (\ref{eq15}), with $w$ running between $-1/3$ at $z\rightarrow\infty$ and $-1$ at $z\rightarrow-1$ for the GB DE and between $-1+\frac{2\gamma}{3\delta}$ at $z\rightarrow\infty$ and $-1$ at $z\rightarrow-1$ for the MGB DE. In absence of matter, according to (\ref{eq7}) the GB DE density at early times exhibits a behavior as $a^{-2}$, which was discussed in \cite{li}, but the energy balance is dominated by the matter behavior $a^{-3}$ for small $a$. The MGB model in absence of matter, and under some relation between the parameters reproduces a $\Lambda$CDM-like model.
The presence of matter drastically changes the behavior of the EoS, allowing the crossing of the phantom barrier at current and late times, presenting a minimum in the future bellow the phantom barrier, and then evolving toward the de sitter phase at $z\rightarrow -1$. This is due to the different nature of the Friedmann equation with the addition of matter content, that converts the Friedmann equation into a nonlinear differential equation. This could be interpreted as follows: the presence of dark matter enforces the effect of the DE at late times, but at early times is clear the dominance of the matter sector (especially in the MGB model), as seen in Figs. 1, 2 and in the statefinder diagrams given in Figs. 5 and 6. A remarkable aspect of this proposal is that independently of the values given to the parameters involved in the model, in all cases the universe evolves toward de Sitter phase in the far future ($z\rightarrow -1$).
Numerical calculations show that despite the fact that the EoS evolves toward the phantom region and ends in the de Sitter phase, the energy density and pressure are free of any finite time future singularities (i.e. $H$ and $\dot{H}$ are finite even in the region of phantom behavior shown in Figs. 1 and 2). This model also makes a concrete prediction about the final state of the universe, namely the universe ends in a de Sitter phase. We performed the statefinder diagnostic and have found the evolution paths in the $s-r$ and $q-r$ planes for the GB and MGB models. For the MGB model the $r(s)$ trajectory is a closed loop that starts at the early time limit of the $\Lambda$CDM fixed point and ends at the late time limit of the $\Lambda$CDM fixed point with de Sitter expansion.
It turns out that this choice of dark energy density (that scales as $L^{-4}$), which can be interpreted as a non-saturated variant of the holographic model, is a viable model which is able to explain the cosmic coincidence, it also leads to quintom behavior but remaining free of finite time future singularities, and is consistent with current observational data.
\section*{Acknowledgments}
This work was supported by Universidad del Valle under project CI 7890.

\end{document}